Full paper

# Conversion of single-energy computed tomography to parametric maps of dual-energy computed tomography using convolutional neural network

Short running title: SECT to parametric maps of DECT with CNN

Sangwook Kim[1,2,*], Jimin Lee[1,3,*], Jungye Kim[4], Bitbyeol Kim[5], Chang Heon Choi[5,6,7,8], Seongmoon Jung[1,5,7,8,†]

[1]Department of Nuclear Engineering, Ulsan National Institute of Science and Technology, Ulsan, 44919, Republic of Korea

[2]Department of Medical Biophysics, University of Toronto, Toronto, Ontario, Canada

[3]Graduate School of Artificial Intelligence, Ulsan National Institute of Science and Technology, Ulsan, 44919, Republic of Korea

[4]Department of Biomedical Engineering, Korea University, Seoul, 02841, Republic of Korea

[5]Department of Radiation Oncology, Seoul National University Hospital, Seoul, 03080, Republic of Korea

[6]Department of Radiation Oncology, Seoul National University College of Medicine, Seoul, 03080, Republic of Korea

[7]Biomedical Research Institute, Seoul National University Hospital, Seoul, 03080, Republic of Korea

[8]Institute of Radiation Medicine, Seoul National University Medical Research Center, Seoul, 03080, Republic of Korea

*Authors equally contribute to this work

†Corresponding author's e-mail: smjung@snu.ac.kr

†Corresponding author's address: Seoul National University Hospital, 101 Daehangro, Jongno-Gu, Seoul, 03080, Korea

Tel: 82 2 2072-4160

**Source of funding**

This research was supported by the National Research Foundation of Korea (NRF) funded by the Ministry of Science and ICT (Grant No. NRF-2020R1F1A1073430, NRF-2021R1F1A1057818, and RS-2023-00252216).

**Conflict of interest**

The authors have no relevant conflicts of interest to disclose.

# ABSTRACT

**Objectives:** We propose a deep learning (DL) multi-task learning framework using convolutional neural network (CNN) for a direct conversion of single-energy CT (SECT) to three different parametric maps of dual-energy CT (DECT): Virtual-monochromatic image (VMI), effective atomic number (EAN), and relative electron density (RED).

**Methods:** We propose VMI-Net for conversion of SECT to 70, 120, and 200 keV VMIs. In addition, EAN-Net and RED-Net were also developed to convert SECT to EAN and RED. We trained and validated our model using 67 patients collected between 2019 and 2020. SECT images with 120 kVp acquired by the DECT (IQon spectral CT, Philips) were used as input, while the VMIs, EAN, and RED acquired by the same device were used as target. The performance of the DL framework was evaluated by absolute difference (AD) and relative difference (RD).

**Results:** The VMI-Net converted 120 kVp SECT to the VMIs with AD of 9.02 Hounsfield Unit, and RD of 0.41% compared to the ground truth VMIs. The ADs of the converted EAN and RED were 0.29 and 0.96, respectively, while the RDs were 1.99% and 0.50% for the converted EAN and RED, respectively.

**Conclusions:** SECT images were directly converted to the three parametric maps of DECT (i.e., VMIs, EAN, and RED). By using this model, one can generate the parametric information from SECT images without DECT device. Our model can help investigate the parametric information from SECT retrospectively.

**Advances in knowledge**: Deep learning framework enables converting SECT to various high-quality parametric maps of DECT.

# 1. Introduction

Dual-energy computed tomography (DECT) is a paramount imaging technique developed for probing different materials with various attenuation properties [1]. While single-energy computed tomography (SECT) is still effective in clinical applications, it struggles to differentiate materials accurately and determine their exact attenuation properties. In contrast, DECT has been suggested to overcome this limitation by combining signals from two distinct energy levels. Combining these signals, various parametric maps can be reproduced from the pair of low- and high-energy signals. This nature of DECT enables superior differentiation of materials compared to SECT.

## 1.1. Parametric maps of DECT in clinical scenarios

There are various types of parametric maps in DECT which can be applied in the clinical scenario providing better imaging quality for improving diagnosis. The virtual monochromatic image (VMI) arises as an imaging technique, synthesized through the combination of low- and high-energy signals and the subsequent decomposition of two distinct materials [1]. Studies have proved the clinical usages of VMI for better diagnosis with the improved visibility. A higher iodine contrast-to-noise ratio was achieved while effectively reducing noise levels, as demonstrated by Kalisz *et al*. in their study of VMIs at low energy levels [2]. Nakaura *et al.* showed the usage of VMIs improving the imaging quality of coronary CT angiography [3]. Pomerantz *et al.* demonstrated that optimizing the energy levels to around 65-75 keV in VMIs improve the quality of unenhanced brain CT images [4]. Moreover, VMIs effectively reduce metal artifacts by exploiting energy-dependent attenuation profiles, differentiating metallic from non-metallic structures while minimizing beam hardening and photon starvation artifacts [5].

Effective atomic number (EAN) is used for material differentiation by quantitatively measuring the changes in attenuation in the CT, leading to better understanding of diagnostic imaging. Nakajima *et al.* show EAN to classify non-calcified coronary plaques [6]. Perfusion defects in lungs resulting from pulmonary embolism can be more easily differentiated from normally perfused lung areas when using the EAN map [1]. Li *et al.* show that EAN of necrosis in non-contrast enhanced scans help differentiating malignant and benign necrotic lung lesions [7].

Relative electron density (RED) map is expressed as an electron density ratio relative to water. RED provides information about the electron density values for materials in the imaging, which is helpful especially for radiation treatment planning. DECT offers numerous advantages in the field of radiotherapy. Estimating accurate stopping-power ratios (SPRs) is critical in planning the delivery of proton and heavy ion therapy, since SPR characterizes the interaction of ion beams with tissue. Faller *et al.* utilized EAN and RED from DECT to show a potential to improve predicting the SPR in the particle therapy treatment planning [8]. Bär *et al.* showed that DECT has an advantage over SECT, reducing proton beam range uncertainties by 0.4% in soft tissues [9].

## 1.2. Deep learning in DECT acquisition

The acquisition of DECT is unattainable even with the high demand due to its high cost. Many studies have focused on applying deep learning (DL) to synthesize DECT from SECT to obtain clinical information via multiple parametric maps unique to DECT. Previously suggested DL-based DECT synthesis methods were dependent on the types of DECT, where the number of sources and detectors varied. Zhao *et al.* propose a DL model trained on 16 patients and tested on three patients, to synthesize VMI images from SECT by generating high energy CT from a low energy CT followed by the reconstruction of VMI with energy ranging from 40 to 150 keV where dual-source SOMATOM Definition Flash DECT scanner (Siemens Healthineers, Forchheim, Germany) was used [10]. Kawahara *et al.* developed a DL-based reconstruction of bone-water and fat-water images from 120 kVp SECT with a Revolution DECT scanner (GE Healthcare, Princeton, NJ, USA) using fast kV switching method [11]. Unlike the two types of DECT scanners, Su *et al.* developed and evaluated various machine learning methods for synthesizing DECT maps acquired from a single-source dual-layer detector: IQon Spectral CT (Philips Healthcare, Amsterdam, Netherlands). They targeted generating EAN, RED, and the SPR from SECT using phantoms and 10 clinically acquired patient datasets. Another study by Lyu *et al.* developed a U-Net based convolutional neural network (CNN) algorithm for estimating DECT using non-contrast-enhanced brain VMIs, trained and tested using 22 patients. This study aimed to train DL model for synthesizing VMIs across different energy levels (70 keV to 110 keV and 150 keV) [12]. Studies showing deep learning-based DECT synthesis are listed and summarized in TABLE 1.

TABLE 1    Summary of studies for deep learning-based synthesis of DECT. We mainly focus on the types of scanner, inputs and outputs for the deep learning model, type of deep learning model, number of subjects, and anatomical regions.

| Study | Scanner | Inputs / Outputs | Methods | Number of Subjects | Anatomical regions |
|---|---|---|---|---|---|
| Zhao et al.[10] | SOMATOM Definition Flash dual-source DECT Scanner (Siemens Healthineers) | Inputs: Low-energy CT (100 kV) | U-Net | 16 | Abdomen |
| | | Outputs: (i) High-energy CT (140 kV) (ii) VMI (40~150 KeV) | | | |
| Kawahara et al. [11] | Revolution DECT Scanner (GE Healthcare) | Inputs: SECT (120 kVp) | DCGAN | 28 | Chest |
| | | Outputs: Material decomposition images: bone and fat | | | |
| Su et al.[13] | IQon Spectral CT (Philips Healthcare) | Inputs: VMIs (70 KeV and 140 KeV), Photoelectric image, and Compton image | Historical centroid, Random forest, Artificial neural network | 10 | Abdomen |
| | | Outputs: EAN, RED, Mean excitation energy, and relative stopping power | | | |
| Lyu et al. [12] | IQon Spectral CT (Philips Healthcare) | Inputs: Low-energy VMI (70 and 110 KeV) | U-Net | 22 | Brain |
| | | Outputs: High-energy VMI (110 and 150 KeV) | | | |

SECT (either 100 kVp or 120 kVp) images have been used to generate 140 kVp images, VMIs, and material decomposition images [10,11]. However those CT or parametric images were acquired by DECT devices using the fast kV switching or the dual-source methods. Other studies have shown a potential of DL-based synthesis of single-source dual-layer DECT scans. However, those are limited in converting parametric maps of DECT to other parametric maps of DECT within a specific target region [12,13]. Furthermore, their models have been developed and validated using small amounts of datasets. In this article, we propose a DL model that convert SECT to various DECT parametric maps (i.e., 70 keV, 120 keV, and 200 keV VMIs, EAN, and RED) using single-source dual-layer DECT for multiple anatomical sites (i.e., head and neck, chest, and abdomen). In addition, we conducted training and testing on both contrast-enhanced and non-contrast-enhanced DECT scans at once, aligning with the goal of creating robust DL models that can be applied effectively without the need to develop separate models for specific cases.

## 2. Materials

### 2.1. Dataset

The dataset included CT images taken using a 120 kVp X-rays as SECT images from 67 patients using with a dual-layer DECT scanner (IQon Spectral CT, Philips) between 2019 and 2020. Using low- and high-energy images, VMIs (70 keV, 120 keV, and 200 keV), RED, and EAN are generated in its software (IntelliSpace Portal, Philips). When the signals from dual-layer detector panels are not separated and then reconstructed, conventional 120 kVp CT images can be obtained. We denoted such CT dataset as SECT in this study. We collected datasets from various anatomical sites. Out of 67 patients, 19 had chest scans, 23 had abdominal and pelvic scans, and 25 had head and brain scans. We included all SECT images acquired with contrast-enhanced (46 scans) and non-contrast-enhanced (21 scans) in both training and test datasets. For model development, we partitioned the dataset into training, validation, and non-overlapping patient-wise test sets, ensuring the test data remained unseen during training. We used 46, 10, and 11 patients for train, validation, and test datasets, respectively. We employed a 2D CNN model and trained and tested it using 2D axial slices, totaling 12,100 slices, with 9,011, 1,692, and 1,397 in each split. This study was performed in line with the principles of the Declaration of Helsinki. Approval was granted by the Institutional Review Board (IRB approval no. 2111-198-1280). Written informed consent by the patients was waived due to a retrospective nature of our study.

### 2.2. Preprocessing

The preprocessing step is comprised of two steps, min-max intensity normalization and data augmentation. As we implemented our proposed network as a 2D network, we transformed all 3-dimensional CT images into 2D axial slices. We applied min-max intensity normalization for all SECT and DECT maps, making intensity range of all images with the minimum and maximum value of 0 and 1. For SECT, we adjusted the range by adding 1,024 and then normalized pixel values by dividing them by the respective maximum values: 4,095 for 120 kVp SECT and VMIs, 2,308 for EAN, and 4,049 for RED.

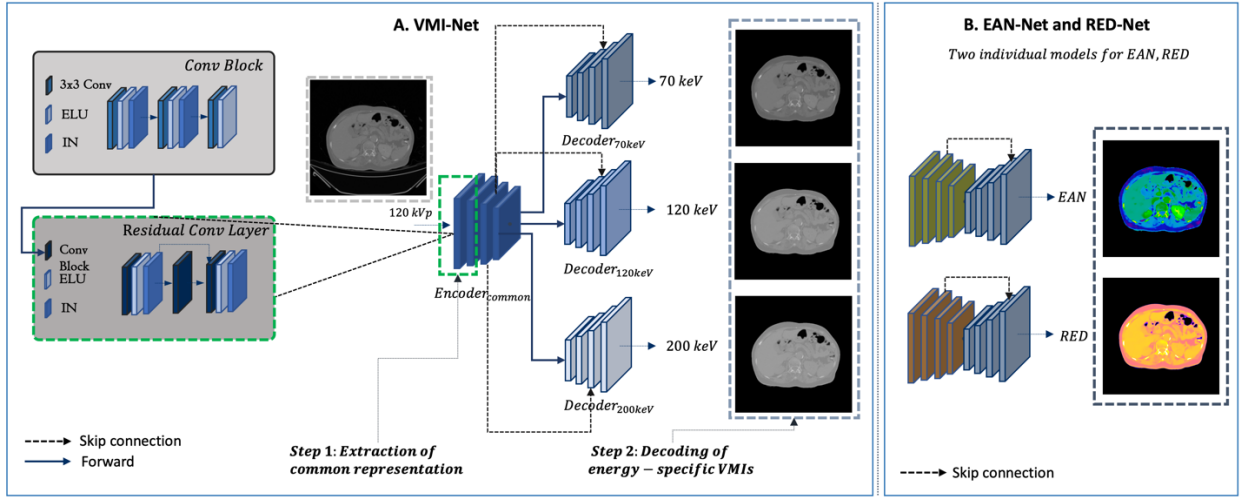

FIGURE 1 Overview of the model development of our proposed approach. (A) VMI-Net with a single encoder and three energy-specific decoders for generating VMIs with three different virtual monochromatic images (VMI) with 70keV, 120keV, and 200keV, simultaneously. IN refers to instance normalization layer. (B) EAN-Net and RED-Net for converting 120 kVp SECT to effective atomic number (EAN) and relative electron density (RED) using two separate encoders and decoders. Best viewed in color.

## 3. Methods

### 3.1. Model development

We implemented our proposed CNN models based on FusionNet [14] which is a fully CNN widely used to solve semantic segmentation and image translation tasks for biomedical images [15]. However, training separate DL models for each VMI is inefficient and impractical due to the numerous potential VMIs with varying energy levels, which would necessitate a corresponding increase in the number of required models. In this regard, we propose using multi-task learning, which can simultaneously synthesize three VMIs by jointly training to convert SECT. As shown in FIGURE 1A, VMI-Net has a single encoder followed by multiple decoders convert SECT to three different VMIs. This model efficiently uses shared encoder representation to generate VMIs, leveraging the inherent linear relationship between input SECT and the target VMIs.

As shown in FIGURE 1, the backbone encoders and decoders are the same for all models. The input of the model is a single-channel SECT scan. The encoder has four down-sampling blocks with the channel size of 32, 64, 128, and 256, respectively. Residual convolution blocks exist in each down-sampling convolution block, which have convolution block with kernel size of 3, followed by the instance normalization and ELU activation function. Inside the residual convolution block, residual skip-connection connects the first and last blocks in the residual convolution block to prevent gradient vanishing by skipping the gradient across the convolution block at the center. Decoders have four up-sampling blocks, and the structure of blocks are equal to the encoder blocks. The number of channels for up-sampling blocks are symmetric to that of down-sampling blocks, i.e., 256, 128, 64, and 32. Instead of down-sampling, however, decoder blocks have up-sampling blocks using a linear interpolation to up-sample the feature maps from the previous decoder block. The size of the features in the decoder blocks match the size of the feature maps from the encoder, which are copied and summed with the feature maps from the encoder in the skip connection. The final layer of the network generates a single-channel DECT scan.

We implemented two separate FusionNet-based models to convert SECT to EAN and RED as shown in FIGURE 1B, which are EAN-Net and RED-Net, respectively. The network architectures of these two models resemble VMI-Net, except they have a single encoder and decoder, since the model is targeted for generating EAN and RED separately. Synthesis of EAN and RED has been a difficult problem due to the difficulty of capturing the relationship between SECT and EAN and RED. Additionally, unlike VMI-Net, we implemented the two networks for EAN and RED, since both modalities share less commonalities, compared to that of three VMIs.

## 3.2. Evaluation

We evaluated the performance of generated DECT using four metrics to measure the similarity of given images: Pearson correlation coefficient (PCC), peak signal-to-noise ratio (PSNR), structural similarity index (SSIM), and mean squared error (MSE) (see Supplementary material for detail). These metrics determine the pixel-level difference between the DL generated DECT and the ground truth (GT) DECT. It is important to note that only the pixels within the body are considered for the calculation. We also measured absolute difference (AD) and relative difference (RD) which are the difference between the GT and DL-predicted DECT. AD [Hounsfield unit (HU)] and RD (%) are defined as:

$$AD\ (HU) = \frac{\sum_{i=1}^{N}\left|x_{pred}^{i} - x_{gt}^{i}\right|}{N} \qquad (1),$$

$$RD\ (\%) = \frac{\sum_{i=1}^{N} 100 * \frac{\left|x_{pred}^{i} - x_{gt}^{i}\right|}{max\ (x_{pred},\ x_{gt})}}{N} \qquad (2),$$

where $x_{pred}^i$ and $x_{gt}^i$ denote the $i^{th}$ pixel value of DL-predicted output and GT DECT images. $N$ refers to the total number of pixels in the image. The $\max(x_{pred}, x_{gt})$ is the maximum value of the two images: $x_{pred}$ $and$ $x_{gt}$. We report both metrics to better interpret the difference and to avoid misinterpretation of results.

## 4. Results

### 4.1. Conversion of SECT to VMIs

As shown in TABLE 2, VMI-Net showed the average AD of 9.0151 HU and RD of 0.4076%, showing less than 0.5% difference compared to GT VMIs. Quality of synthesized VMIs differs across low to high-energy VMIs, showing that converting SECT to 70 KeV is superior than that of 200 KeV. FIGURE 2 and FIGURE 3 show qualitative results of converting SECT to VMIs for both contrast-enhanced and non-contrast-enhanced VMIs for abdomen scans. Synthesized contrast-enhanced VMIs show comparatively higher difference in blood vessel and kidneys where contrast enhancement is noticeable, while differences for the same regions are lower in non-contrast-enhanced VMIs. The results for head and neck and chest scans were described in FIGURE S1-S4 in Supplementary material.

TABLE 2   Quantitative results of VMI-Net generating virtual monochromatic images from single energy computed tomography 120 kVp images. The mean and standard deviation are calculated across all images in the test datasets. The mean and standard deviation are calculated across all images in the test datasets.

| Metrics | 70 KeV | 120 KeV | 200 KeV | Average |
|---------|--------|---------|---------|---------|
| PCC | 0.9983±0.0002 | 0.9982±0.0003 | 0.9982±0.0005 | 0.9982 ±0.0003 |
| MSE | 80.1063±200.9955 | 96.3575±222.1282 | 128.1929±386.6525 | 101.5522±269.9254 |
| PSNR | 54.8822±2.8404 | 54.1266±2.8211 | 53.4783±3.0961 | 54.1624±2.9192 |
| SSIM | 0.9997±0.0004 | 0.9996±0.0003 | 0.9996±0.0004 | 0.9996±0.0004 |
| AD (HU) | 8.8414±1.8858 | 8.8957±1.6002 | 9.3082±1.6712 | 9.0151±1.7191 |
| RD (%) | 0.3378±0.0762 | 0.4194±0.0624 | 0.4656±0.0582 | 0.4076±0.0656 |

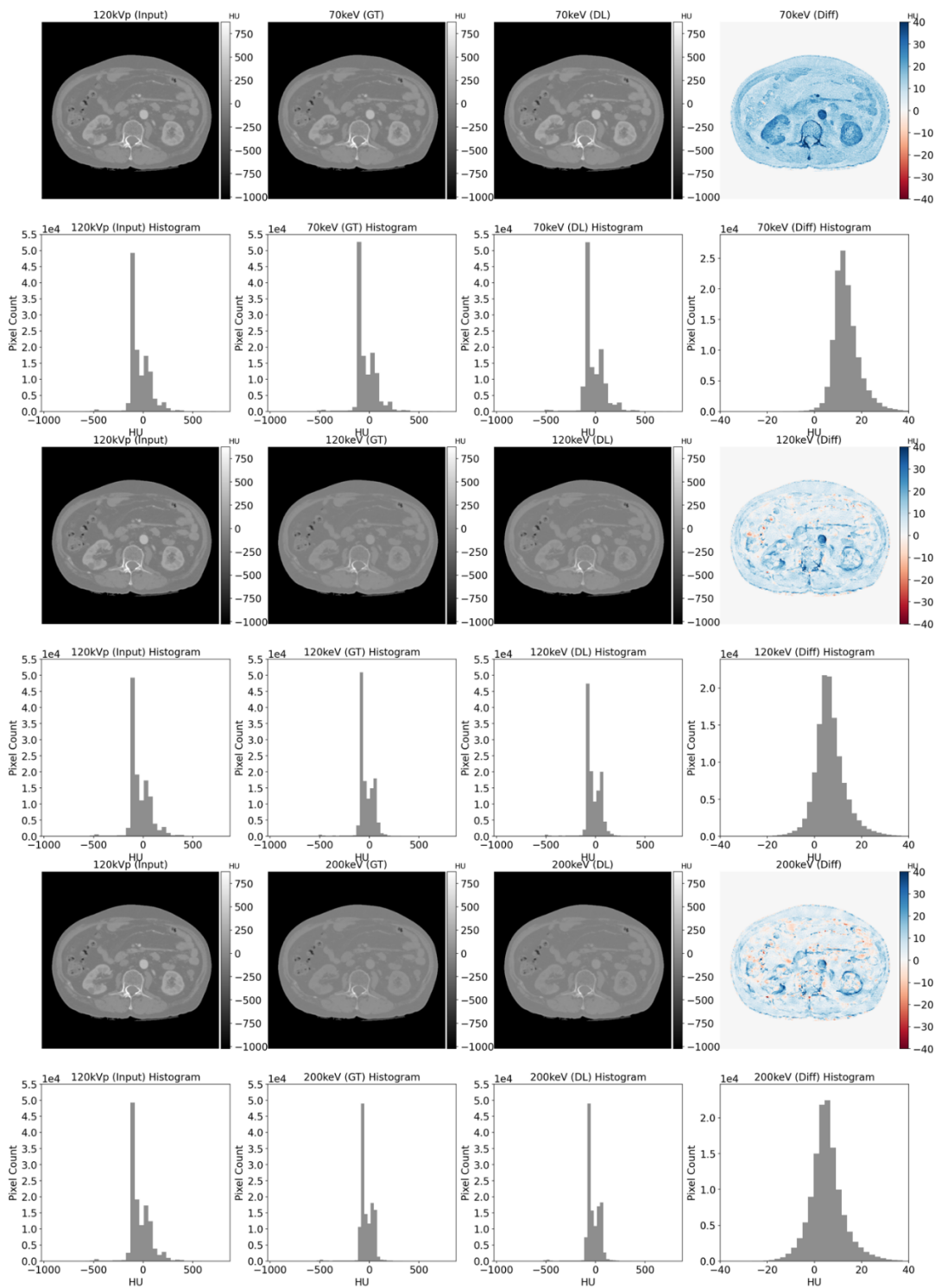

FIGURE 2 Qualitative results of VMI-Net for contrast-enhanced VMIs. From the left to the right columns, input 120 kVp single energy computed tomography (SECT) for abdomen, ground truth (GT) virtual monochromatic images (VMIs), generated VMIs, and the difference between the GT and the generated VMIs of 70, 120, and 200 keV. The histogram of each image is shown under each image.

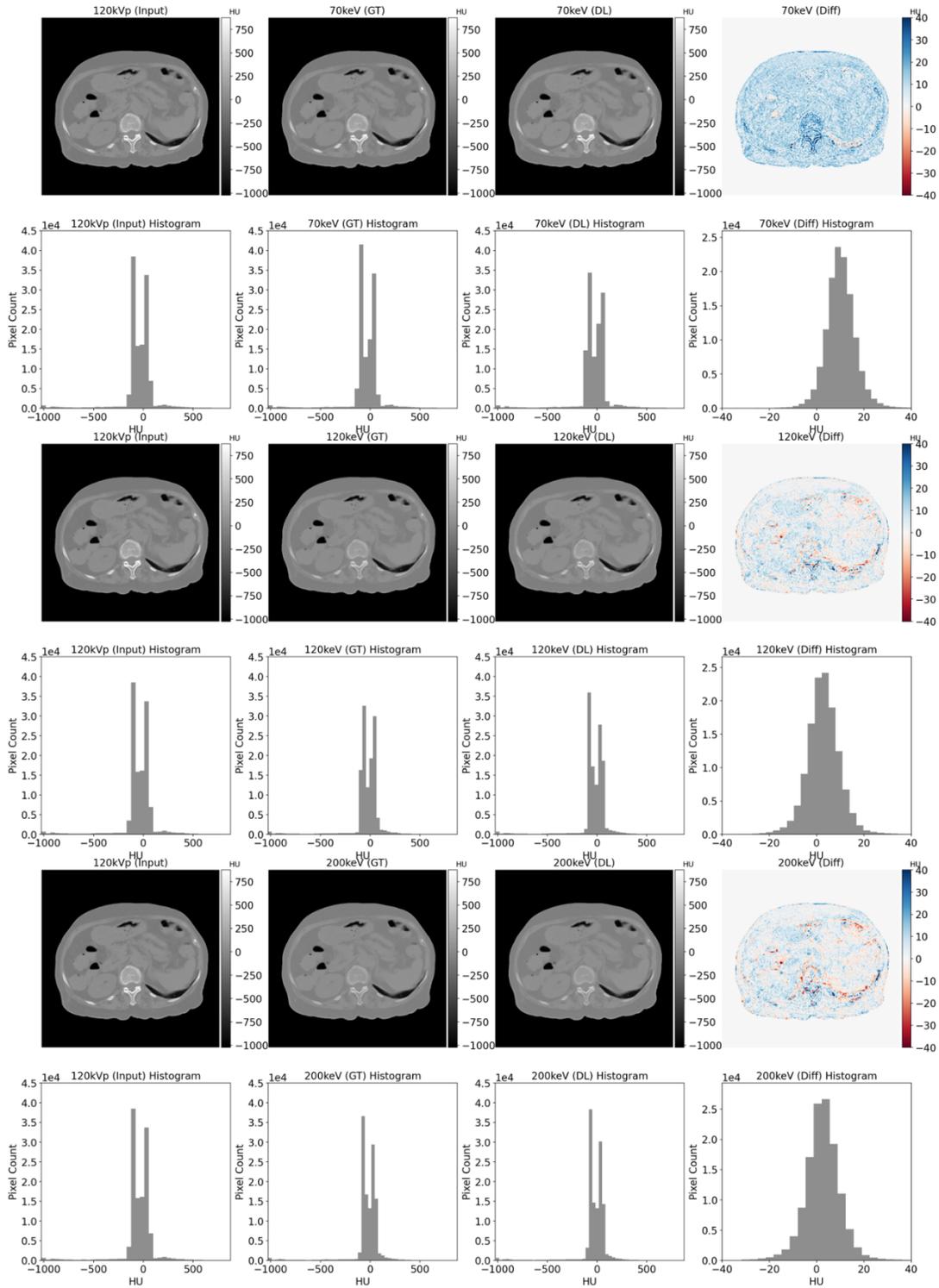

FIGURE 3    Qualitative results of VMI-Net for non-contrast-enhanced VMI generation. From the left to the right columns, input 120 kVp single energy computed tomography (SECT) for abdomen, ground truth (GT) virtual monochromatic images (VMIs), generated VMIs, and the difference between the GT and the generated VMIs of 70, 120, and 200 keV. The histogram of each image is shown under each image.

### 4.2. Conversion of SECT to EAN and RED

In TABLE 3, the conversion performance of RED shows comparatively higher than that of EAN. RED generation showed a RD of 0.50%, while generating EAN showed 2.05%, demonstrating EAN generation is much harder than RED since SECT and RED has relatively stronger relationship compared to that of SECT and EAN. FIGURE 4 and 5 illustrate qualitative results of converting contrast-enhanced and non-contrast-enhanced SECT to EAN, respectively. Our proposed approach effectively translated structural information from SECT but exhibited mediocre performance in predicting exact atomic numbers within organs and generating textures, mainly due to limited correlations between SECT and EAN compared to other DECT maps. FIGURE 6 and 7 show qualitative results of RED synthesis from contrast and non-contrast-enhanced SECT. Results show that DL-generated EAN shows larger variance compared to VMIs or RED. We provide an overview of the findings of the research study. FIGURE 8 represents the mean and standard deviation of RD between the GT and DL generated DECT. The results for head and neck and chest scans were described in FIGURE S5-S9 in Supplementary material.

### 4.3. Performance of Contrast-enhanced and Non-contrast-enhanced DECT Synthesis

In TABLE 4, we evaluated the performance of VMIs, EAN, and RED synthesis using contrast-enhanced and non-contrast-enhanced SECT scans. Out of 11 test datasets, six are contrast-enhanced scans, while five are non-contrast-enhanced images. We compared both using RD and we showed the average of three VMIs. Again, FIGURE 2, 4, 6 show qualitative results for converting contrast-enhanced VMIs, EAN, and RED while FIGURE 3, 5, 7 illustrate generated DECTs from non-contrast-enhanced SECT. Overall, both qualitative and quantitative results show trivial difference, showing that our models are robust regardless of contrast-enhancement.

TABLE 3    Quantitative results of generating effective atomic number (EAN) and relative electron density (RED) using respective models for each parametric map.

| Metrics | EAN | RED |
| --- | --- | --- |
| PCC | 0.9951±0.0028 | 0.9981±0.0007 |
| MSE | 0.1116±0.0968 | 1.5114±4.9473 |
| PSNR | 40.4600±3.5671 | 52.9946±3.1861 |
| SSIM | 0.9943±0.0043 | 0.9995±0.0005 |
| AD | 0.2925±0.0705 | 0.9640±0.1723 |
| RD (%) | 1.9854±0.5506 | 0.5011±0.0529 |

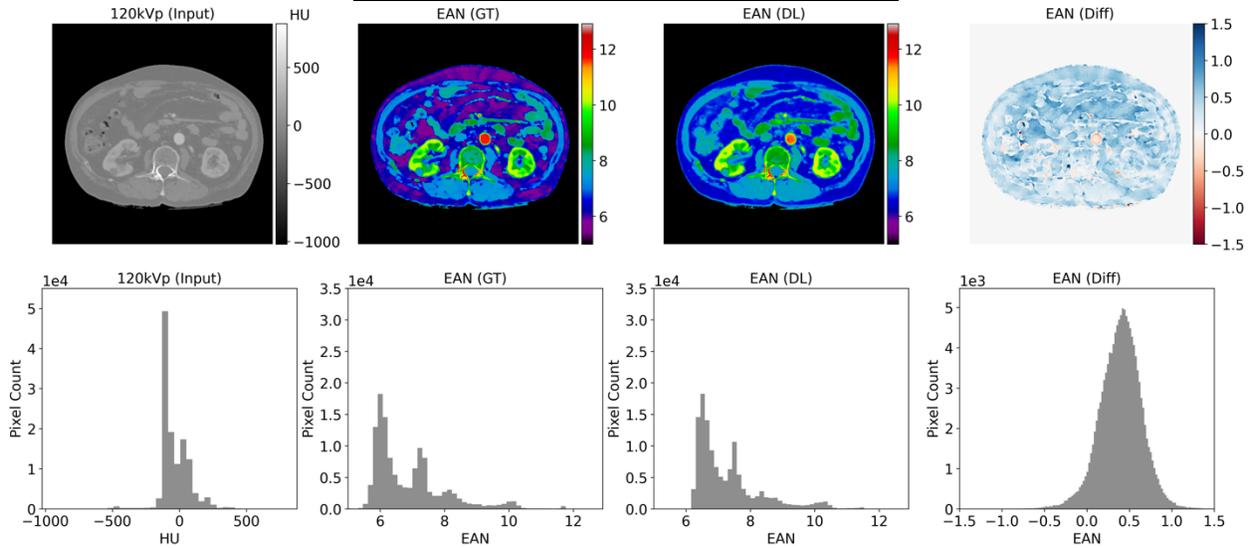

FIGURE 4    Qualitative results of converting contrast-enhanced single energy computed tomography (SECT) to effective atomic number (EAN). From the left to the right, input SECT, ground truth (GT) EAN, generated EAN, and the difference between the GT and generated EAN.

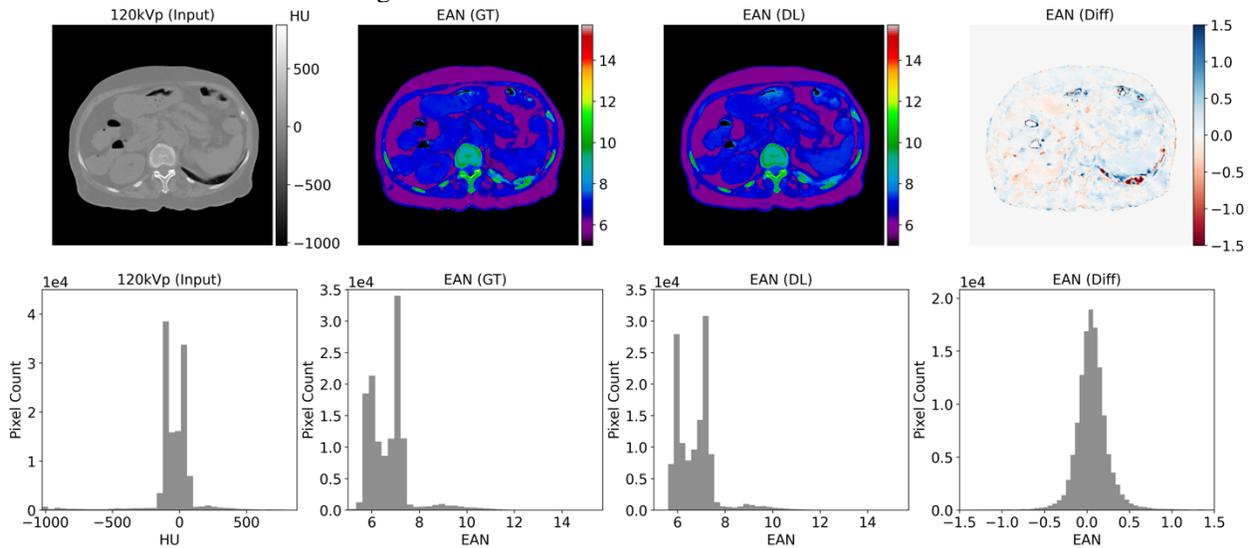

FIGURE 5    Qualitative results of converting **non-contrast-enhanced** single energy computed tomography (SECT) to effective atomic number (EAN). From the left to the right, input SECT, ground truth (GT) EAN, generated EAN, and the difference between the GT and generated EAN.

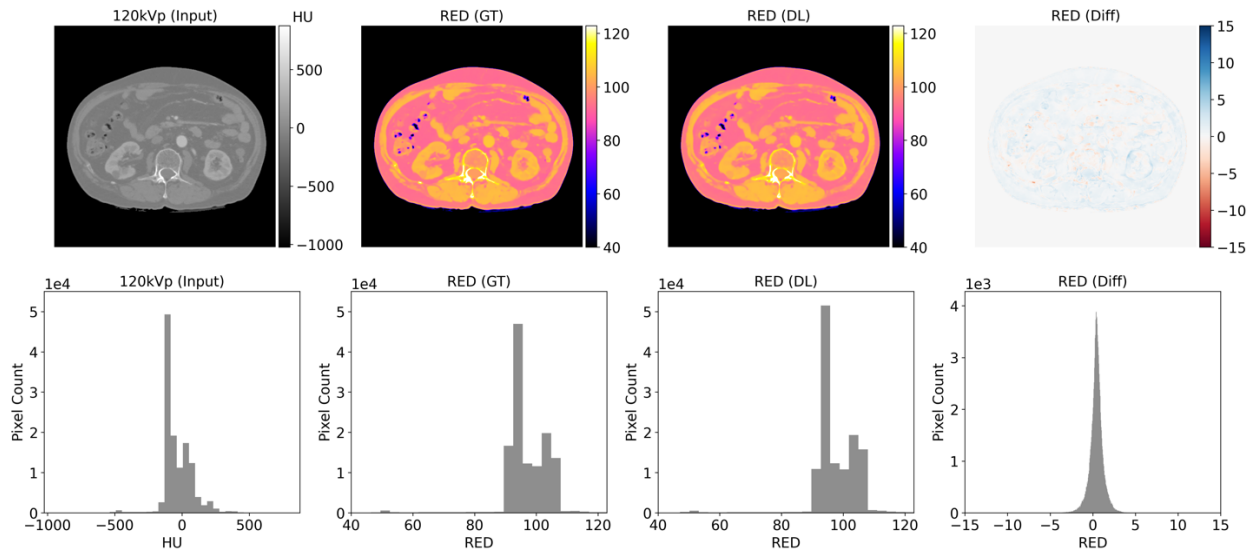

FIGURE 6  Qualitative results of converting contrast-enhanced single energy computed tomography (SECT) to relative electron density (RED). From the left to the right, input SECT, ground truth (GT) RED, generated RED, and the difference between the GT and generated RED.

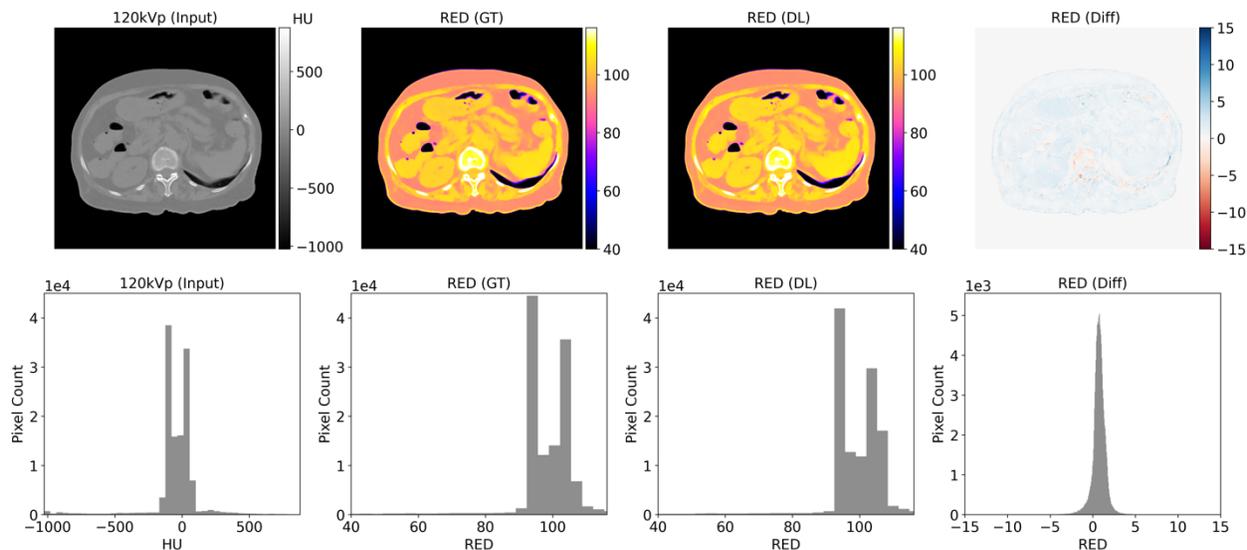

FIGURE 7  Qualitative results of converting non-contrast-enhanced single energy computed tomography (SECT) to relative electron density (RED). From the left to the right, input SECT, ground truth (GT) RED, generated RED, and the difference between the GT and generated RED.

TABLE 4    Quantitative results of synthesizing DECT from SECT using contrast and non-contrast SECT scans. We compare contrast-enhanced and non-contrast-enhanced test cases using six evaluation metrics (i.e., Pearson correlation coefficient (PCC), mean squared error (MSE), peak signal-to-noise ratio (PSNR), structural similarity index (SSIM), relative difference (RD) (%), and absolute difference (AD)) for VMIs, EAN, and RED. We averaged three VMIs (70 keV, 120 keV, and 200 keV) to represent the overall performance of VMIs.

| Parametric maps | Metrics | Contrast | Non-Contrast |
|---|---|---|---|
| VMIs (Average) | PCC | 0.9983±0.0003 | 0.9982±0.0002 |
| | MSE | 127.0337±346.5219 | 64.5600±46.1825 |
| | PSNR | 53.7612±3.3740 | 54.7447±1.9424 |
| | SSIM | 0.9997±0.0003 | 0.9995±0.0003 |
| | AD (HU) | 9.8067±1.7522 | 8.0652±1.4705 |
| | RD (%) | 0.3947±0.0646 | 0.4230±0.0771 |
| EAN | PCC | 0.9956±0.0026 | 0.9944±0.0030 |
| | MSE | 0.1110±0.1018 | 0.1125±0.0894 |
| | PSNR | 40.6201±3.7775 | 40.2276±3.2304 |
| | SSIM | 0.9953±0.0032 | 0.9927±0.0051 |
| | AD | 0.3047±0.0725 | 0.2780±0.0813 |
| | RD (%) | 1.9174±0.6219 | 2.0669±0.5788 |
| RED | PCC | 0.9980±0.0009 | 0.9982±0.0002 |
| | MSE | 1.9963±6.3645 | 0.8074±0.6641 |
| | PSNR | 52.4619±3.6649 | 53.7678±2.1004 |
| | SSIM | 0.9995±0.0006 | 0.9995±0.0003 |
| | AD | 1.0760±0.1568 | 0.8296±0.0976 |
| | RD (%) | 0.4863±0.0633 | 0.5189±0.0445 |

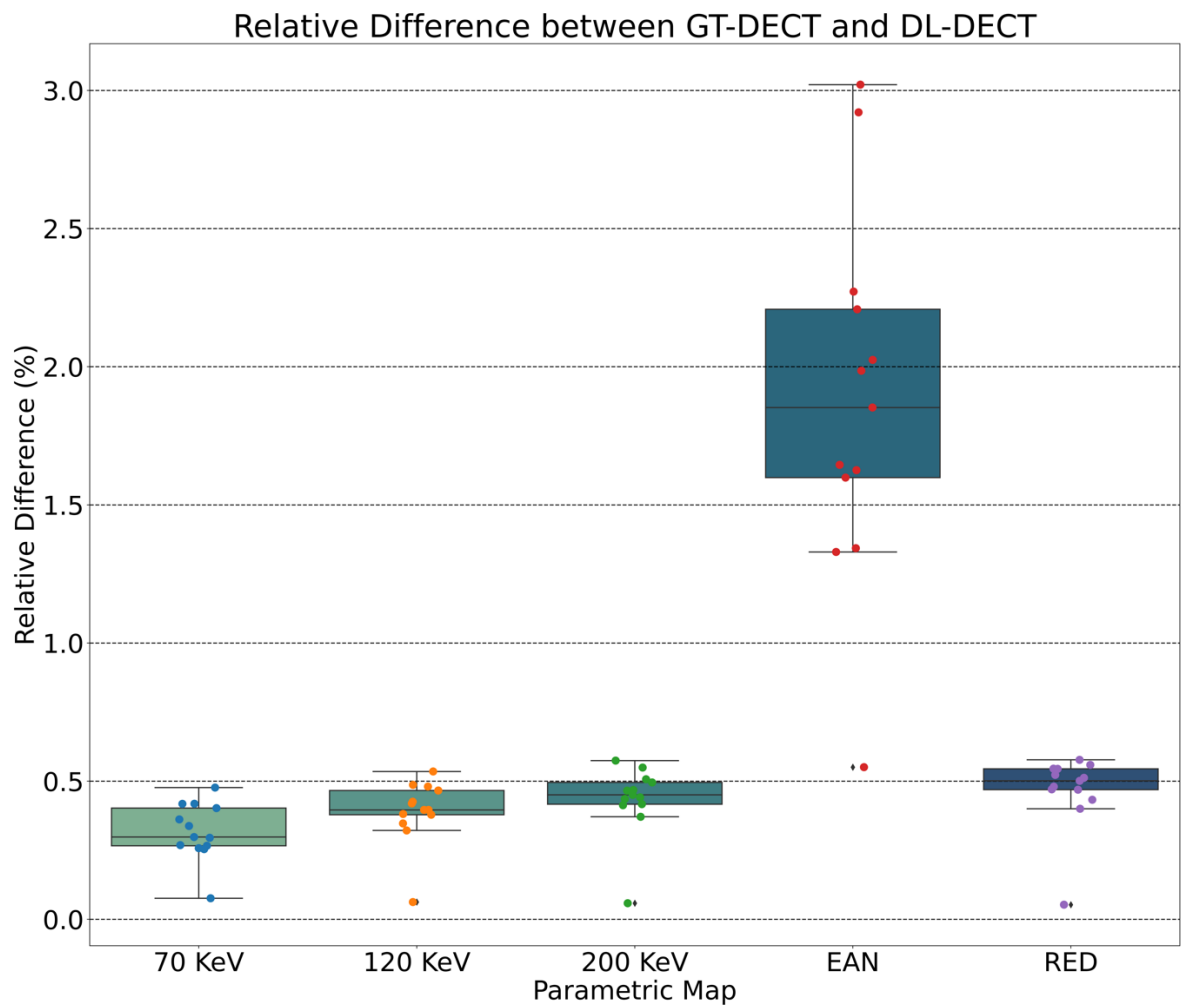

FIGURE 8   Boxplots represent the mean and the standard deviation of relative difference between the ground truth and deep learning generated dual-energy CT scans. Each dot represents sample from the test datasets. Median and the standard deviation of each parametric map are illustrated within the box. The figures are best viewed in color.

## 5. Discussion

We developed a DL-based framework that can effectively convert SECT into various types of parametric maps of DECT with less than 0.5% of relative difference compared to the GT VMIs. By utilizing shared encoder information, the model captures the linear relationship between target VMIs, as they share the same anatomy as the input SECT but differ in attenuation. This shared bottleneck feature enhances robustness against noise in the input SECT.

In practice, DL models often suffer from out-of-distribution dataset, which is not generally seen when training the model. In this regard, training both non-contrast-enhanced and contrast-enhanced DECT with a single model might hurt its performance due to the opposing information between the two. For example, HU values are significantly different for contrast-enhanced soft tissue. In FIGURE 4, the synthesized EAN sample shown in the figure showed RD of 3.10% which is significantly higher than average performance which is 1.92%, showing that estimating enhanced EAN values from contrast-enhanced CT is comparatively challenging. As shown in TABLE 4, we observed that the quantitative performance for synthesizing non-contrast-enhanced DECT slightly outperformed the performance for contrast-enhanced DECT. However, the overall difference between the two is trivial, proving that our model is generalizable regardless of contrast enhancement in the CT scans. This is critical for generalizability of DL model as it reduces burden for training separate models for each DECT map.

As shown in TABLE 3, synthesizing EAN showed inferior performance compared to that of other parametric maps. This can be attributed to the weak relationship between the EAN and attenuation coefficient of SECT. Since CNN operate by capturing the relationship between input and output images, the poor performance of EAN-Net can be explained by the lack of adequate information connecting the HU values of SECT and EAN. To address this, additional information that can establish a stronger relationship between SECT and EAN can be incorporated into the DL model, e.g., using different imaging modalities. This remains as a future direction of our work.

Our proposed methods are valuable in situations where the use of DECT equipment is limited. To the best of our knowledge, this is the first time demonstrating a feasibility of DL model for directly generating DECT from IQON Spectral CT scanner, reconstructing VMI, EAN, and RED along with conventional 120 kVp SECT. Results present the clinical feasibility of our approach which can keep the physical property of GT VMIs. With the high quality synthesized DECT scans from SECT, it can greatly contribute to improving quality diagnostic and treatment planning by providing informative DECT, maintaining their clinical advantages over SECT.

However, the application of this model can be limited due to variations in SECT acquisition across vendors. Therefore, the generalizability of our model is a critical consideration. To address this, we suggest an additional step of SECT conversion from one domain to another before applying our method. Recent studies have shown a potential of deep learning-based conversion across different vendors to improve generalizability of developed deep learning models. Kim *et al*. developed RouteGAN which can convert CT across different vendors [16]. Hwang *et al*. demonstrated clinical utility of RouteGAN to improve the accuracy and variability for interstitial lung disease quantification [17]. To ensure our model's adaptability across varied clinical settings, rigorous generalizability validation is a key focus of our future research.

## 6. Conclusion

In conclusion, our study demonstrates the effectiveness of DL in converting SECT to various DECT parametric maps from multiple anatomical sites. While our framework successfully generated VMIs, EAN and RED images from SECT, the accuracy of EAN generation was limited due to the low correlation between SECT and EAN, which we leave as a future research direction. Overall, the results of our study suggest that the generated DECT maps has a potential to improve diagnosis and treatment planning even without a complex and high-cost DECT scanner, with only using routinely acquired SECT.

**Ethical statement**

This study was performed in line with the principles of the Declaration of Helsinki. Approval was granted by the Institutional Review Board (IRB approval no. 2111-198-1280). Written informed consent by the patients was waived due to a retrospective nature of our study.

**Availability of data and materials**

This study was pre-printed in arXiv. The data that support the findings of this study are available on request from the corresponding author. The data are not publicly available due to their containing information that could compromise the privacy of research participants.

# Conversion of single-energy computed tomography to parametric maps of dual-energy computed tomography using convolutional neural network

Short running title: SECT to parametric maps of DECT with CNN

# 1. Image similarity metrics for quantitative evaluation

### a. Pearson Correlation Coefficient (PCC)

PCC is used to measure disparity, image registration, and distinguish objects between two images. In this study, we use PCC to evaluate how the GT and generated images are similar or dissimilar, which refers to the calculation of disparity between the two images (Eq. 1). We denote GT images and predicted outputs from deep learning models as $x_{gt}$ and $x_{pred}$, respectively. The definition of PCC is defined as:

$$PCC = \frac{\sum_{i=1}^{N}(x_{gt}^i - x_{gt}^m)(x_{pred}^i - x_{pred}^m)}{\sqrt{\sum_{i=1}^{N}(x_{gt}^i - x_{gt}^m)^2}\sqrt{\sum_{i=1}^{N}(x_{pred}^i - x_{pred}^m)^2}} \qquad (1)$$

where $x^i$ refers to the pixel with the index $i$ from the entire image with a total of $N$ pixels, and $x^m$ refers to the mean value of the pixel intensity values throughout the image.

### b. Mean Square Error (MSE)

MSE is a metric that can be used to calculate the distance between pixels from the GT and generated images. We calculate MSE using:

$$MSE = \frac{1}{MN}\sum_{i=1}^{M}\sum_{j=1}^{N}\left(x_{gt}^{ij} - x_{pred}^{ij}\right)^2 \qquad (2)$$

where $i$ and $j$ refers to the pixel index of rows and columns of images with the number of $M$ pixels and $N$ pixels, respectively.

### c. Peak Signal-to-Noise Ratio (PSNR)

PSNR is a general method to evaluate the quality of a given image by measuring the difference between the original image and the given image. PSNR refers to the ratio between the maximum possible value (power) of a signal or image and the distortion caused by noise (Eq. 3). Here, the noise refers to the difference added to the generated image so that we calculate how noisy the generated image is. $R$ refers to the maximum fluctuation of the image type, which is 1 for our study. $MSE$ refers to the mean squared error which is described in Eq. 2.

$$PSNR = 10\log_{10}\frac{R^2}{MSE} \qquad (3)$$

### d. Structural Similarity Index (SSIM)

SSIM is used to measure the similarity between two images. The GT image is used as a reference image so that SSIM calculates how similar the GT and generated images are (Eq. 4). In our experiments, we set the window size of SSIM as 7.

$$SSIM(x, y) = \frac{(2\mu_x\mu_y + c_1)(2\sigma_{xy} + c_2)}{(\mu_x^2 + \mu_y^2 + c_1)(\sigma_x^2 + \sigma_y^2 + c_2)} \qquad (4)$$

Note that $x$ and $y$ refers to the location of the window. $\mu$ and $\sigma$ refers to the mean and standard deviation of the values within the window. We denote $c$ as a constant preventing the zero division.

## 2. Qualitative results for synthesizing DECT

### a. Qualitative results for synthesizing VMIs

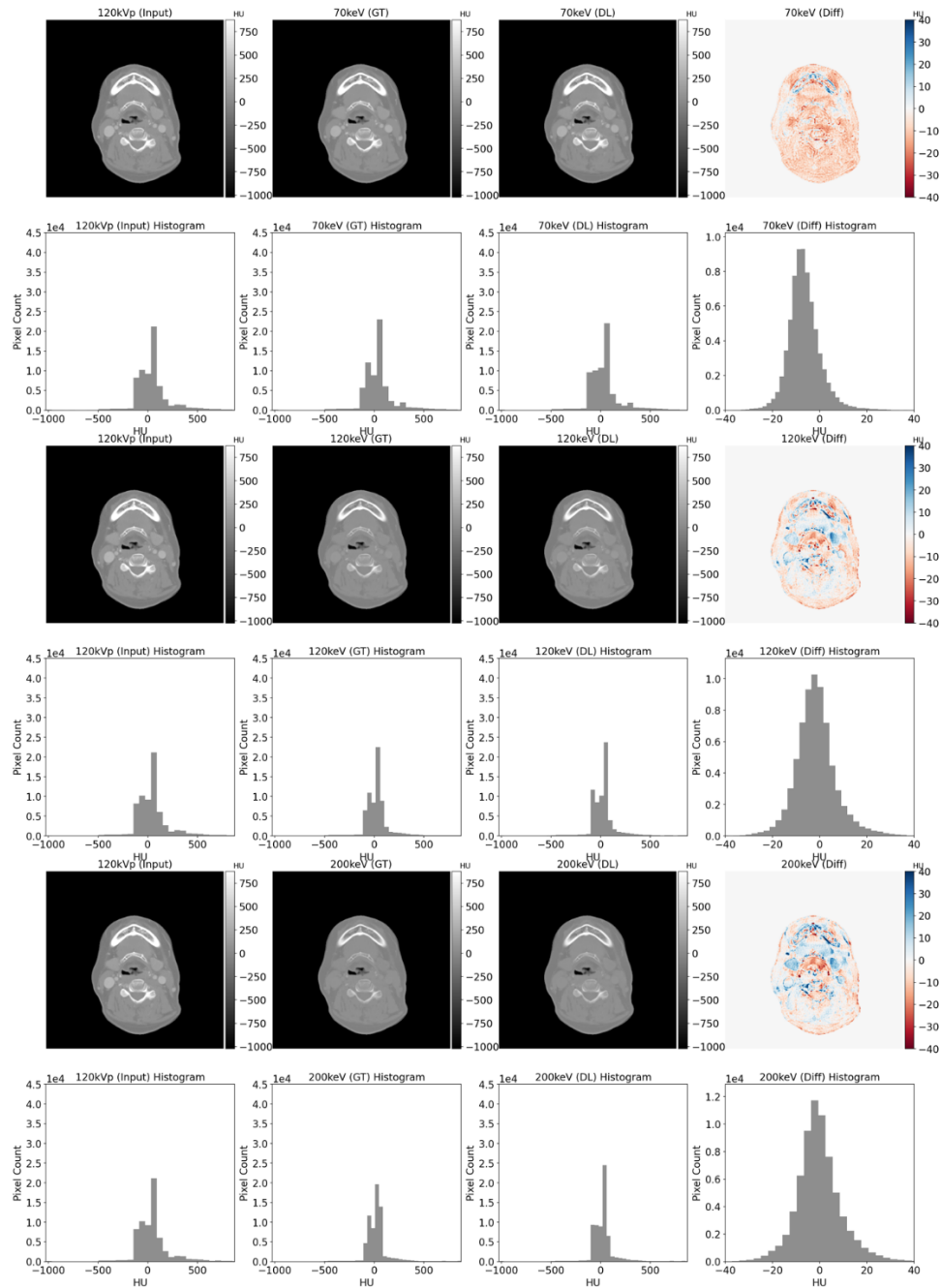

FIGURE S1   Qualitative results of VMI-Net. From the left to the right columns, input contrast-enhanced head and neck 120 kVp single energy computed tomography (SECT), ground truth (GT) virtual monochromatic images (VMIs), generated VMIs, and the difference between the GT and the generated VMIs of 70, 120, and 200 keV. The histogram of each image is shown under each image.

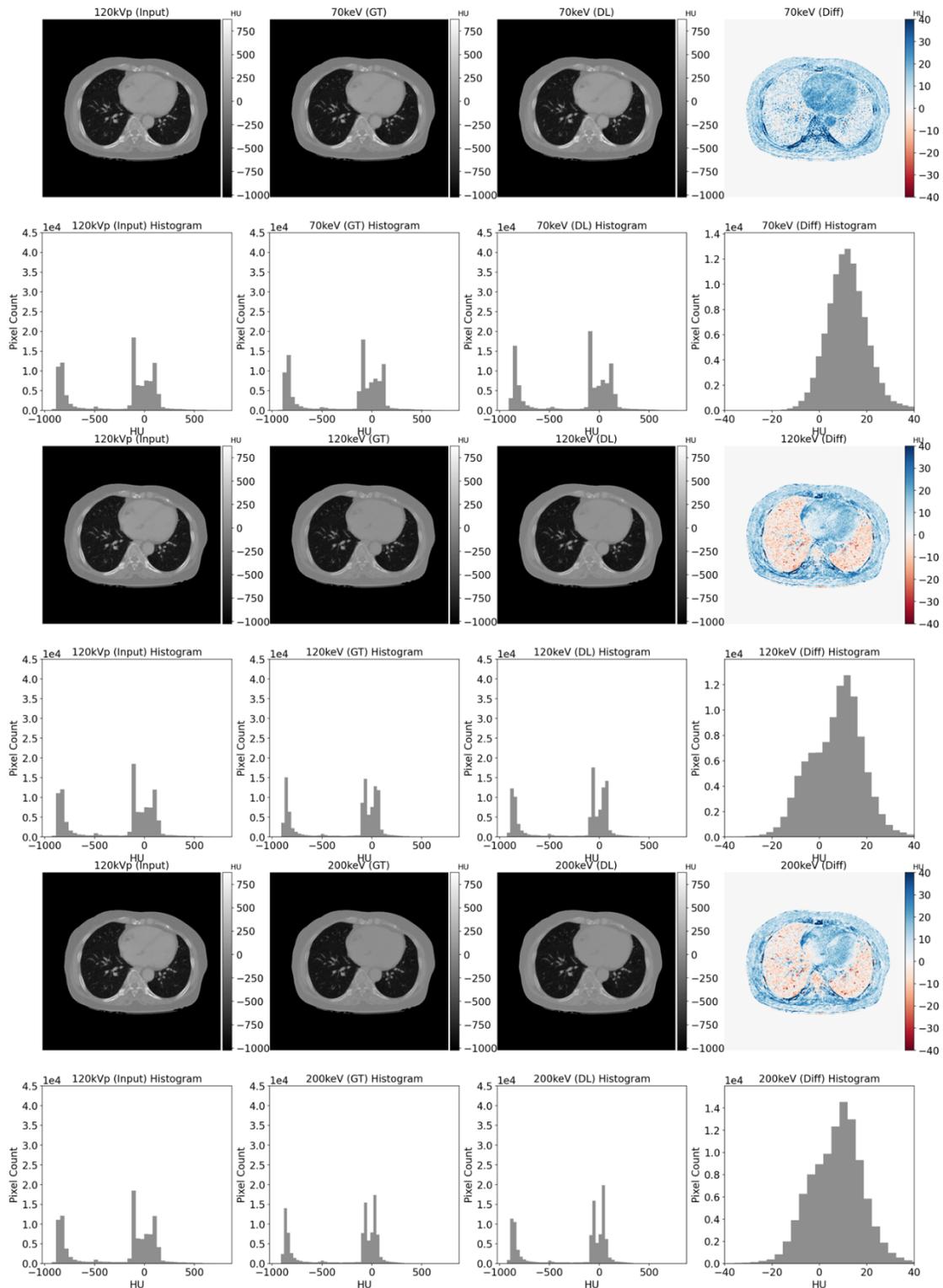

FIGURE S2    Qualitative results of VMI-Net. From the left to the right columns, input contrast-enhanced chest 120 kVp single energy computed tomography (SECT), ground truth (GT) virtual monochromatic images (VMIs), generated VMIs, and the difference between the GT and the generated VMIs of 70, 120, and 200 keV. The histogram of each image is shown under each image.

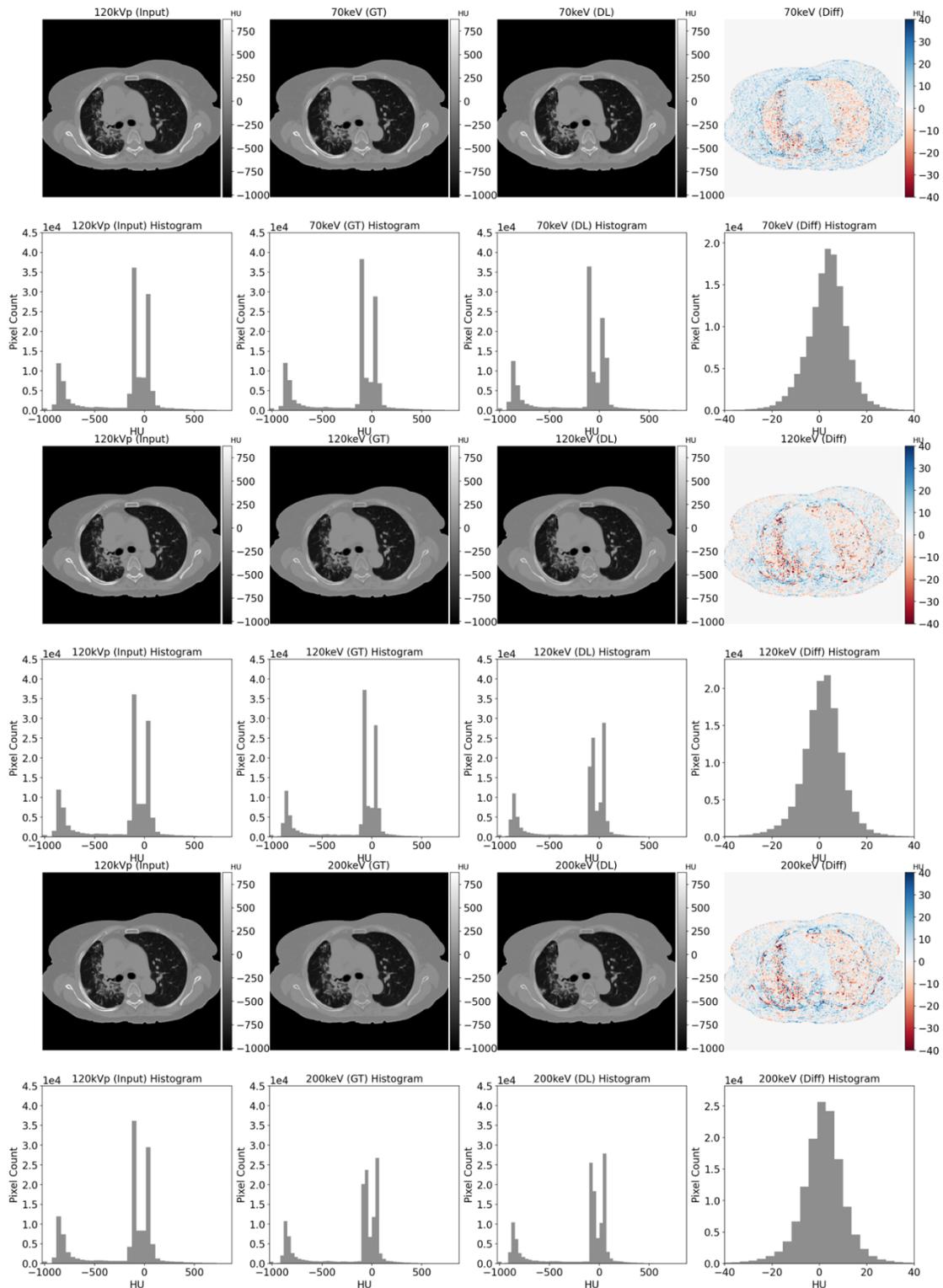

FIGURE S3  Qualitative results of VMI-Net. From the left to the right columns, input non-contrast-enhanced chest 120 kVp single energy computed tomography (SECT), ground truth (GT) virtual monochromatic images (VMIs), generated VMIs, and the difference between the GT and the generated VMIs of 70, 120, and 200 keV. The histogram of each image is shown under each image.

## b. Qualitative results for synthesizing EAN

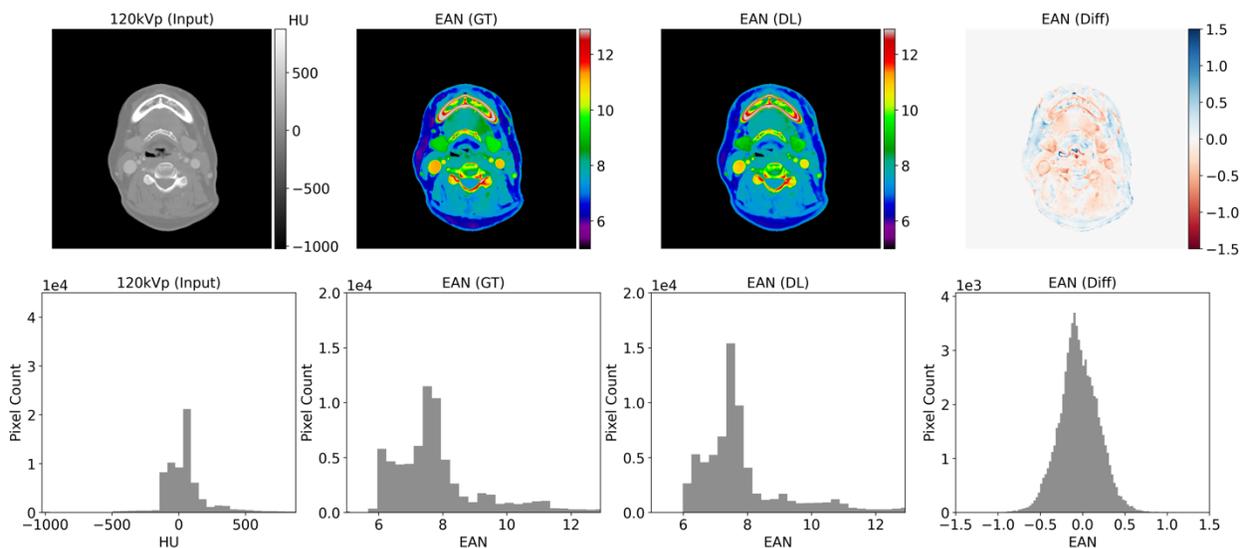

FIGURE S4 Qualitative results of converting contrast-enhanced head and neck single energy computed tomography (SECT) to effective atomic number (EAN). From the left to the right, input SECT, ground truth (GT) EAN, generated EAN, and the difference between the GT and generated EAN.

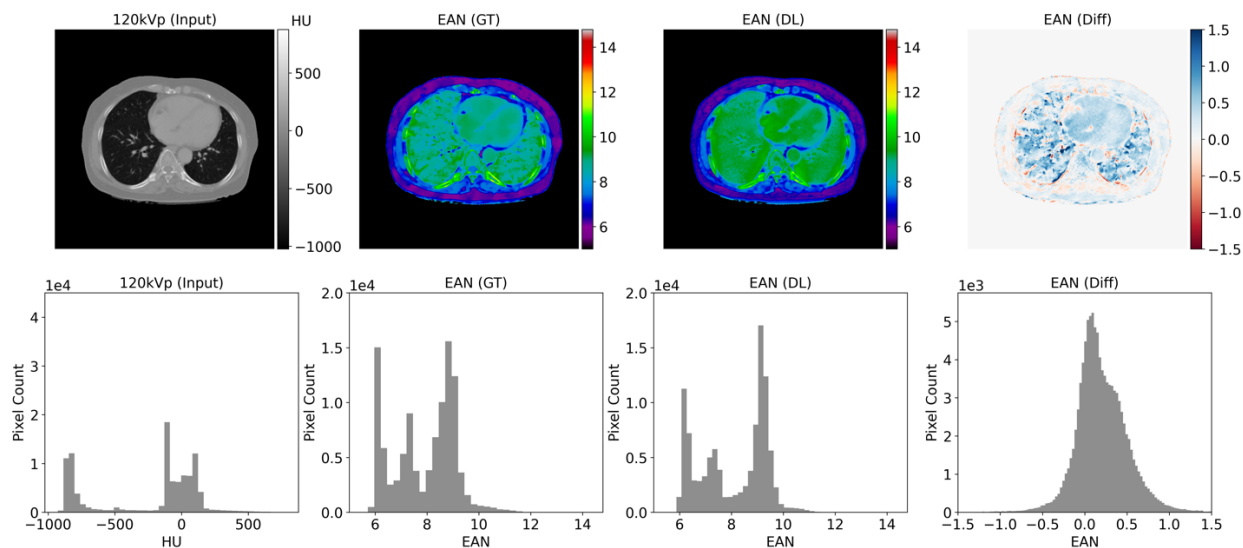

FIGURE S5 Qualitative results of converting contrast-enhanced chest single energy computed tomography (SECT) to effective atomic number (EAN). From the left to the right, input SECT, ground truth (GT) EAN, generated EAN, and the difference between the GT and generated EAN.

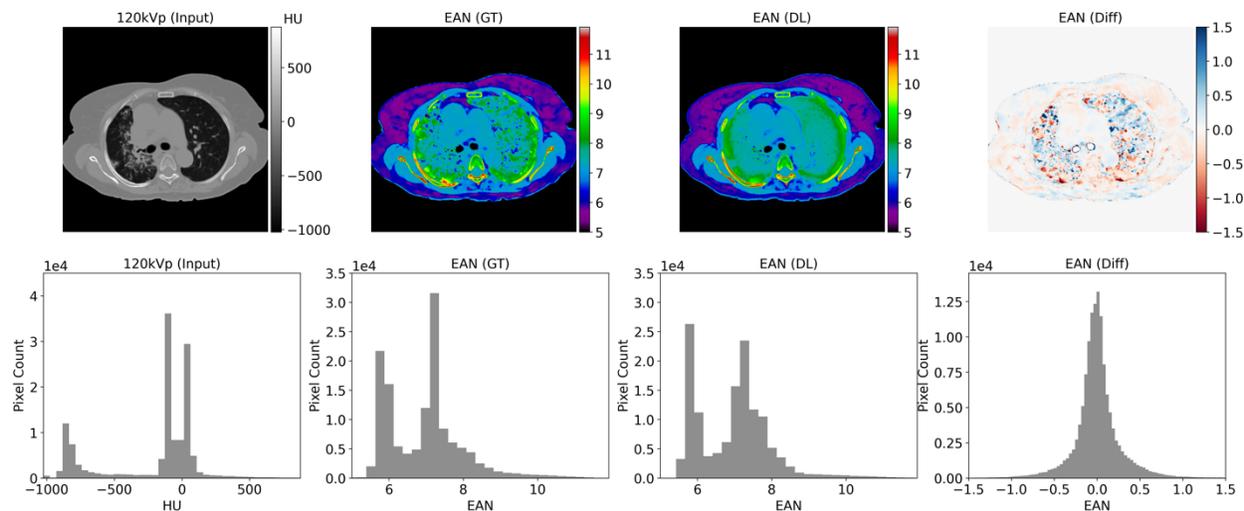

FIGURE S7   Qualitative results of converting non-contrast-enhanced chest single energy computed tomography (SECT) to effective atomic number (EAN). From the left to the right, input SECT, ground truth (GT) EAN, generated EAN, and the difference between the GT and generated EAN.

## c. Qualitative results for synthesizing RED

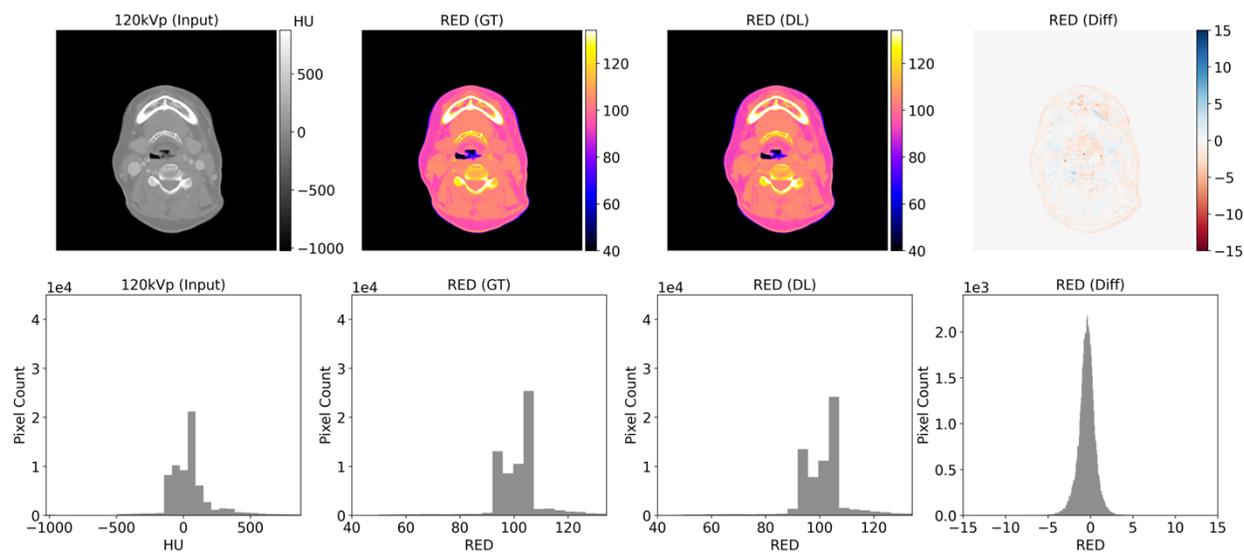

FIGURE S7   Qualitative results of converting contrast-enhanced head and neck single energy computed tomography (SECT) to relative electron density (RED). From the left to the right, input SECT, ground truth (GT) RED, generated RED, and the difference between the GT and generated RED.

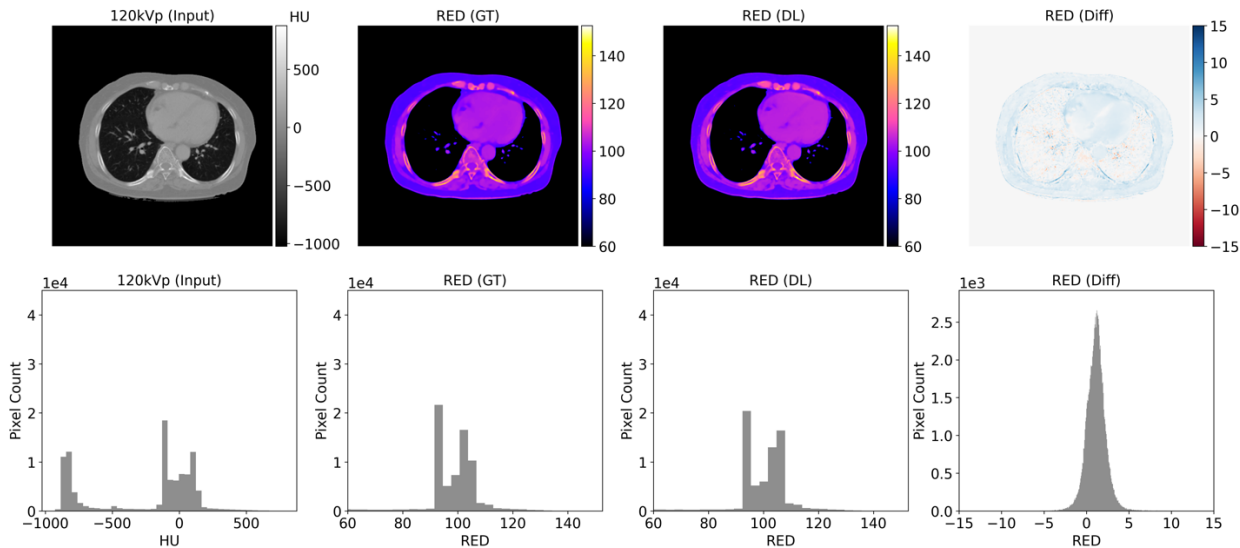

FIGURE S8   Qualitative results of converting contrast-enhanced chest single energy computed tomography (SECT) to relative electron density (RED). From the left to the right, input SECT, ground truth (GT) RED, generated RED, and the difference between the GT and generated RED.

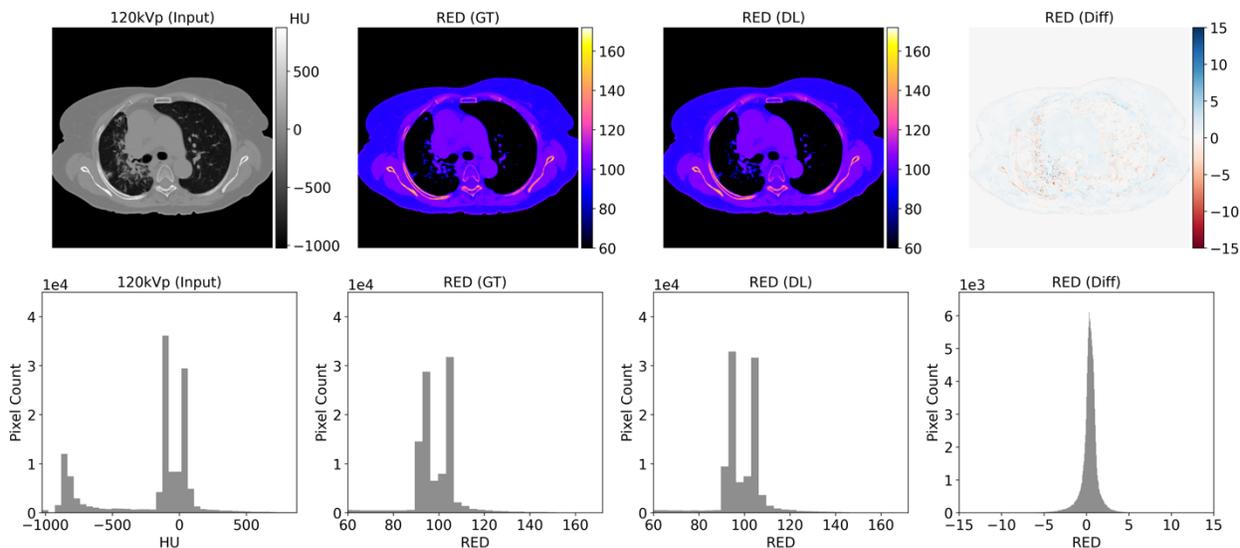

FIGURE S9   Qualitative results of converting non-contrast-enhanced chest single energy computed tomography (SECT) to relative electron density (RED). From the left to the right, input SECT, ground truth (GT) RED, generated RED, and the difference between the GT and generated RED.